\documentclass[a4paper, amsfonts, amssymb, amsmath, reprint, showkeys, nofootinbib, twoside]{revtex4-1}
\usepackage[english]{babel}
\usepackage{tabularx}
\usepackage[utf8]{inputenc}
\usepackage[colorinlistoftodos, color=green!40, prependcaption]{todonotes}
\usepackage[pdftex, pdftitle={Article}, pdfauthor={Author}]{hyperref} 
\usepackage[version=4]{mhchem}
\usepackage{hyperref}
\usepackage{gensymb}
\begin{document}

\title{Chemical tuning of electronic and transport properties of the Bi-Se-Te family of topological insulators}
\author{Maxwell Doyle$^1$, Benjamin Schrunk$^{1, 2}$, D. L. Schlagel$^2$, Thomas A. Lograsso$^2$, Adam Kaminski$^{1, 2}$}
  \affiliation{$^1$Department of Physics and Astronomy, Iowa State University, Ames, Iowa 50011, USA}
  \affiliation{$^2$Division of Materials Science and Engineering, Ames Laboratory,
US DOE, Ames, Iowa 50011, USA. }

\date{\today} 

\begin{abstract}
We use laser-based Angle-Resolved Photoemission Spectroscopy (ARPES) to study how chemical substitution modifies the electronic properties of the Bi$_2$(Se$_{1-x}$Te$_x$)$_3$ (BiSeTe) family of topological insulators. We find that increasing the Te content lowers the chemical potential, leading to a decrease in the binding energy of the Dirac point and a reduction in the density of states originating from the bulk band. This reduction leads to a transition from metallic to semiconducting temperature dependence of the resistivity. For the highest Te concentration, the resistivity nearly saturates at the lowest temperatures. The presence of this plateau indicates that metallic topological surface states dominate the conductance, opening the possibility of studying their transport properties.

\end{abstract}

\maketitle

\section{\label{sec:level1}Introduction\protect}
Topological insulators are a class of materials characterized by a bulk insulating gap and the emergence of metallic states at the surface, driven by the nontrivial topology of their electronic bands. These systems are classified as quantum spin Hall states and do not break time-reversal symmetry. \cite{Qi_2010,PhysRevLett.98.106803} The nontrivial spin structure found in the surface states of topological insulators is expected to be helpful in advancing spintronics. \cite{Vobornik2011,Xiu2011}

\indent Topological insulators such as \ce{Bi2Te3}, \ce{Bi2Se3}, and \ce{Sb2Te3} have been predicted and subsequently studied using ARPES. \cite{Zhang2009,Locatelli_2022,Noh_2008} These studies revealed many interesting aspects of topological insulators, such as spin-polarized bands and spin-textured Fermi surfaces. Most of these stoichiometric materials share the same weakness: the presence of a bulk band close to or crossing the chemical potential. These states dominate the transport properties, making it difficult to study the properties of the surface states. This problem can potentially be addressed by expanding the search to intermediate materials, such as the non-stoichiometric BiSeTe system \cite{10.1063/1.4871280,Kushwaha2016}, where the chemical potential can be tuned by adjusting the composition of the samples. This process introduces impurity scattering; however, the surface states are intrinsically resistant to non-magnetic scattering. Unlike trivial materials, where each state is doubly occupied by electrons with opposing spins, surface states in topological insulators exhibit single spin occupancy, and electrons at opposite momentum values ($k$ and $-k$) have opposite spins. \cite{RevModPhys.83.1057,doi:10.7566/JPSJ.82.102001} This suppresses scattering by non-magnetic impurities, as electrons cannot scatter into a state with opposite momentum without a spin-flip process. \cite{Roushan2009} An example of such resilience is a study demonstrating that the surface states of topological insulators survive brief exposure of the sample surface to air. \cite{doi:10.1073/pnas.1115555109,PhysRevB.86.085112} At the same time, prolonged exposure to low-pressure hydrogen has been shown to substantially change electron filling and shift the chemical potential. \cite{PhysRevB.86.085112} By introducing magnetic elements to break time-reversal symmetry \cite{Chang_Wei_Moodera_2014}, these materials exhibit the quantum anomalous Hall effect and provide a versatile platform for the study of Majorana fermions. \cite{10.1063/5.0039059,Klimovskikh2025}

Understanding how topological insulators behave in ambient environments is crucial for their implementation in quantum computing architectures. Here, we present results demonstrating how chemical substitution can be used to tune the band structure and enhance the contribution of topological effects to the resistivity in these materials.

\section{\label{sec:level2}Experimental Details\protect}
Single crystals were grown using appropriate ratios of high-purity bismuth (99.999\%), selenium (99.999\%), and tellurium (99.999\%). The elements were sealed in a quartz tube with a pointed tip and melted into an ingot in an induction furnace to homogenize the composition. The ingot was then sealed in a larger-diameter quartz tube and loaded into a Bridgman furnace. Crystals were grown by withdrawing the quartz tube at a rate of 1 mm/hr after heating to $800^\circ$C.\

\indent ARPES measurements were performed using a laboratory-based system with a vacuum ultraviolet (VUV) laser ARPES spectrometer consisting of a Scienta DA30 electron analyzer and a tunable picosecond Ti:sapphire oscillator coupled to a fourth-harmonic generator. The measurements were carried out using a photon wavelength of 745 nm and an energy of 6.6 eV. The angular resolution was 0.2° and 0.1° perpendicular and parallel to the analyzer slit, respectively. The energy resolution was set to 1 meV. The samples were mounted on copper pucks and cleaved \textit{in situ} at 40 K under a pressure of $4 \times 10^{-11}$ Torr, producing flat, specular surfaces.

\indent The concentration levels of each element were determined by energy-dispersive X-ray spectroscopy (EDS) using a Thermo NORAN Microanalysis System (model C10001) attached to a JEOL scanning electron microscope (SEM). An acceleration voltage of 21 kV, a working distance of 10 mm, and a take-off angle of 35° were used to measure all crystals with unknown composition. These concentrations were used in Figure 2 when plotting quantities as a function of composition. The composition data were also used to verify the stoichiometry of the BiSeTe samples.

\section{\label{sec:level3} Results and Discussion\protect}

The ARPES data of the Fermi surfaces and band dispersions of Bi$2$(Se${1-x}$Te$_x$)$_3$ samples for three different Te concentrations are shown in Figure 1. Panels (a–c) show that the shape of the Fermi surface evolves from hexagonal to more rounded with increasing Te concentration. The band dispersion along the $\Gamma$ cut is shown in panels (d–f). Panel (d) is particularly interesting because the projection of the bulk conduction band is visible below the chemical potential. \cite{PhysRevB.86.085112} We observe that the binding energy of the Dirac point decreases due to the lowering of $E_F$, which is caused by a decrease in the number of electrons per lattice site with increasing Te substitution. The reduced intensity in the lower-left quadrant, most prominent in panel (a), is due to ARPES matrix elements.

\begin{figure}
    \includegraphics[scale=0.48]{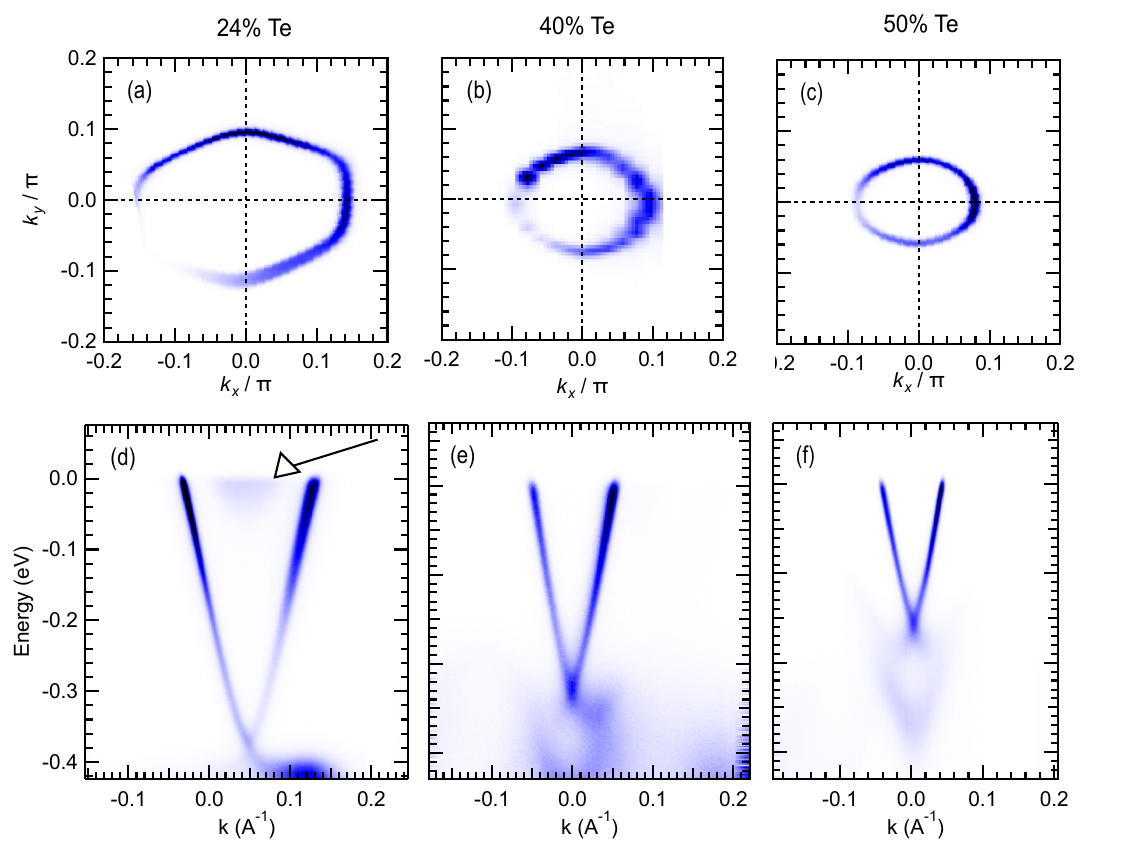}
    \caption{Fermi surface (a-c) and band dispersion (d-f) of BiSeTe for few Te concentrations as indicated above panels. An arrow in panel (d) points to intensity from the bulk band for the lowest Te concentration. }
    \label{fig:placeholder}
\end{figure}
\indent In Figure 2, we plot the energy distribution curves (EDCs) and the Dirac point energy as a function of chemical composition. The peaks indicate the energy positions of the Dirac points and allow us to determine their exact locations and their dependence on substitution, as discussed above. The additional peak near the chemical potential, seen in panel (a) in the green EDC, is due to the presence of a conduction band, as explained above. In panel (b), we quantify the dependence of the Dirac point binding energy on chemical composition by plotting the peak energies from panel (a) as a function of Te concentration. We note that this dependence is approximately linear. The changes in the size of the Fermi surface are shown in panel (c) as a function of tellurium content. We observe a decrease in $\Delta k$ with increasing Te concentration, which is consistent with the data shown in panel (b).\\
\begin{figure}
    \includegraphics[scale=0.42]{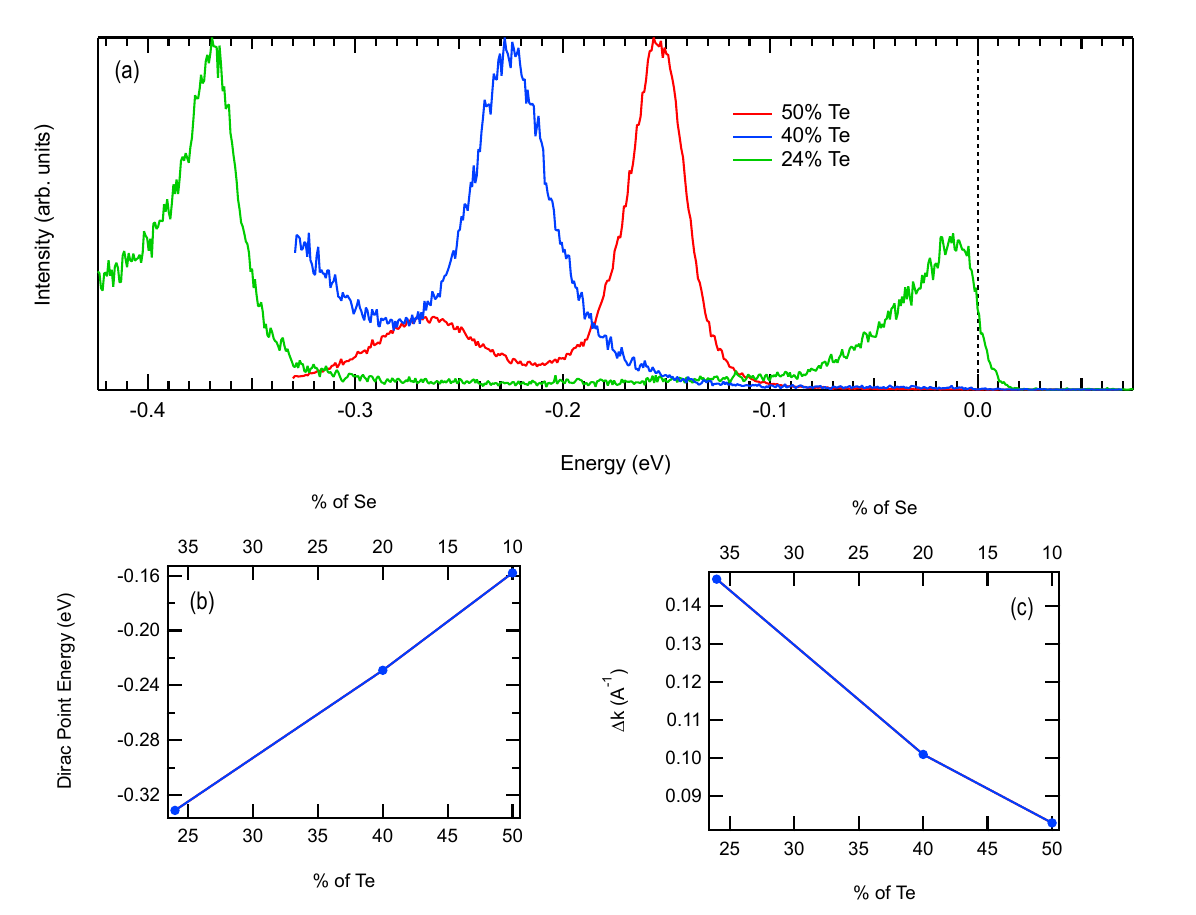}
    \caption{(a) Energy distribution curves showing the decrease of binding energy of Dirac point with increasing Te concentration. 
    (b) Energy of the Dirac point as a function of Te content extracted by fitting peak positions in (a). (c) Fermi caliper size extracted from Fig. 1. }
\end{figure}
\begin{table}
    \centering
    \caption{EDS measurements of chemical composition  of BiSeTe samples.}
    \begin{tabularx}{0.47 \textwidth}{rrr|rrr|rrr}
    \hline
        \multicolumn{3}{c}{Low Te}  &  \multicolumn{3}{c}{Med Te} & \multicolumn{3}{c}{High Te}  \\
        \hline
        ~~~~Bi & ~~~~Se & ~~~~Te & ~~~~Bi & ~~~~Se & ~~~~Te & ~~~~Bi & ~~~~Se & ~~~~Te\\
        \hline
        41.5 & 35.6 & 22.9 & 38.9 & 22.1 & 39.0 & 35.5 & 8.64 & 55.82\\
        40.6 & 34.8 & 24.6 & 40.1 & 20.3 & 39.6 & 37.4 & 7.34 & 55.33\\
        39.8 & 34.8 & 25.5 & 36.5 & 19.2 & 44.4 & 36.0 & 10.55 & 53.44\\
        40.5 & 36.5 & 23.1 & 39.9 & 19.8 & 40.4 & 38.6 & 15.14 & 46.25\\
        39.8 & 37.2 & 23.0 & 39.4 & 19.3 & 41.3 & 36.7 & 10.60 & 52.73\\
        40.9 & 33.7 & 25.4 & 39.8 & 20.6 & 39.6 & 42.3 & 9.34 & 48.34\\
        41.1 & 39.5 & 19.4 & 41.0 & 19.6 & 39.5 & 39.5 & 9.70 & 50.77\\
        40.3 & 37.9 & 21.8 & 39.9 & 20.6 & 39.5 & 37.6 & 11.87 & 50.51\\
        \hline
    \end{tabularx}
    \label{tab:placeholder}
\end{table}
\indent Table 1 displays the results of EDS measurements for different samples. Each column shows the concentration percentages of each element for samples from the low-, medium-, and high-concentration growths, obtained from several measurements on different samples. We note that in Bridgman growths, a concentration gradient from top to bottom is expected. The magnitude of this gradient depends on the travel speed in the heat zone and the length of the ingot. Although the samples used in this study were cut from the center of each ingot, there are still some variations in the Te concentration within each growth. Specifically, in samples with high Te content, we observe Te concentrations ranging from as low as 46\% to as high as nearly 56\%.\

\indent For consistency, we show in Fig. 3 ARPES data measured on the same cleaved surface for each Te concentration at different locations spanning the maximum variation in the Dirac point energy. We again observe a clear increase in the Dirac point energy with increasing Te concentration. We also note that although there are small differences in the Dirac point energy due to slight inhomogeneity in the Te content, these variations are much smaller than the differences between the three Te concentrations. This assures us that the data presented here are representative.

\begin{figure}
   \includegraphics[scale=0.44]{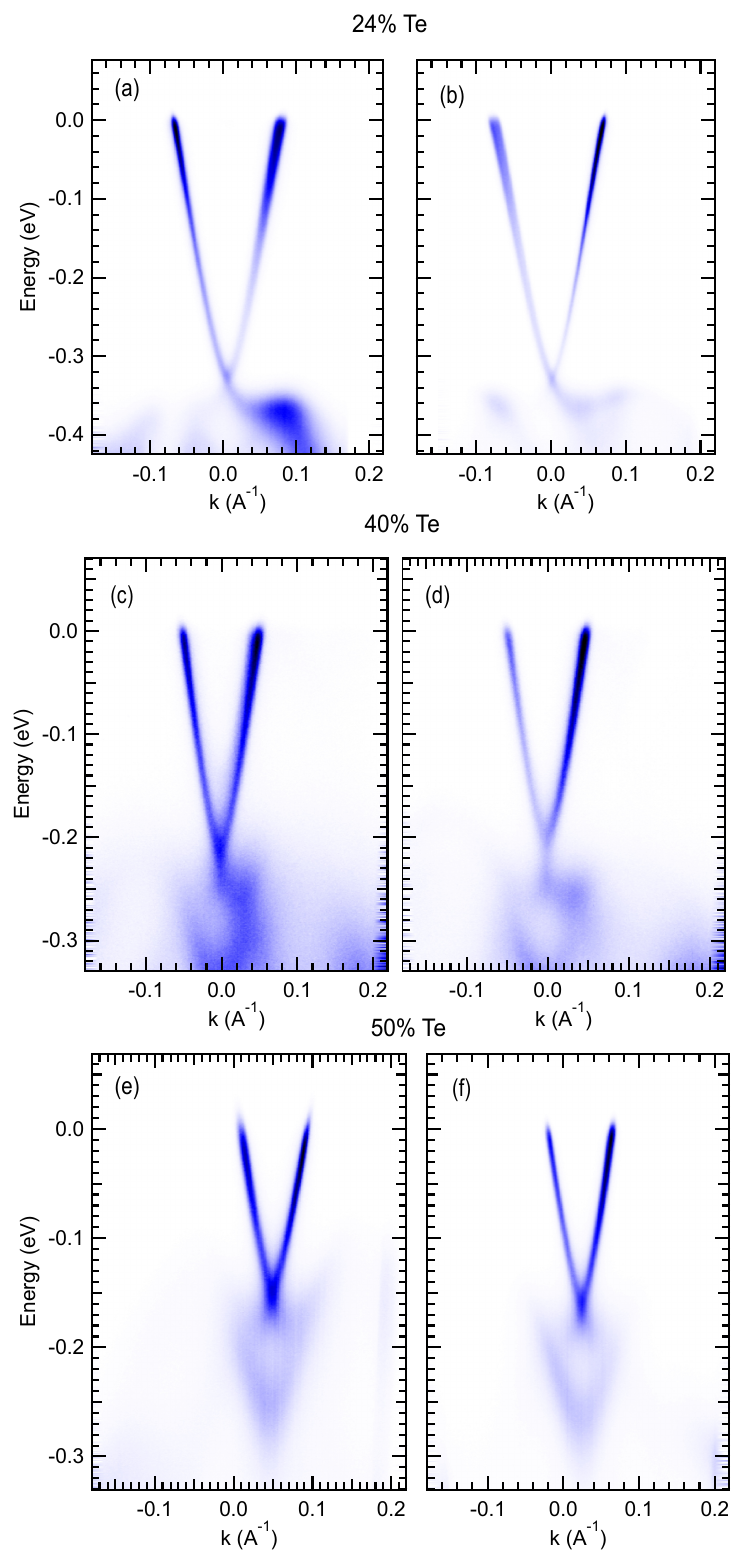} 
   \caption{Illustration of small differences in band dispersion for crystals from the same batch. 
   a-b) for 24~\% Te content, 
   c-d) for 40~\% Te content, 
   e-f) for 50~\% Te content} 
\end{figure}
\indent Similar to the EDC plots shown in Fig. 2, we can plot momentum distribution curves (MDCs) to extract qualitative information about the size of the Fermi surface. To compare different concentration levels, we plot the MDC peaks in the left and right halves of the Brillouin zone in Figure 4, panels (a) and (b), respectively. We observe that the low-Te-concentration MDC has the largest $\Delta k$, and $\Delta k$ decreases with increasing Te concentration. We also note that although the peak positions shift monotonically, the peak widths exhibit interesting behavior; namely, some of the peaks are significantly broader than others.\\
\begin{figure}
    \includegraphics[scale=0.48]{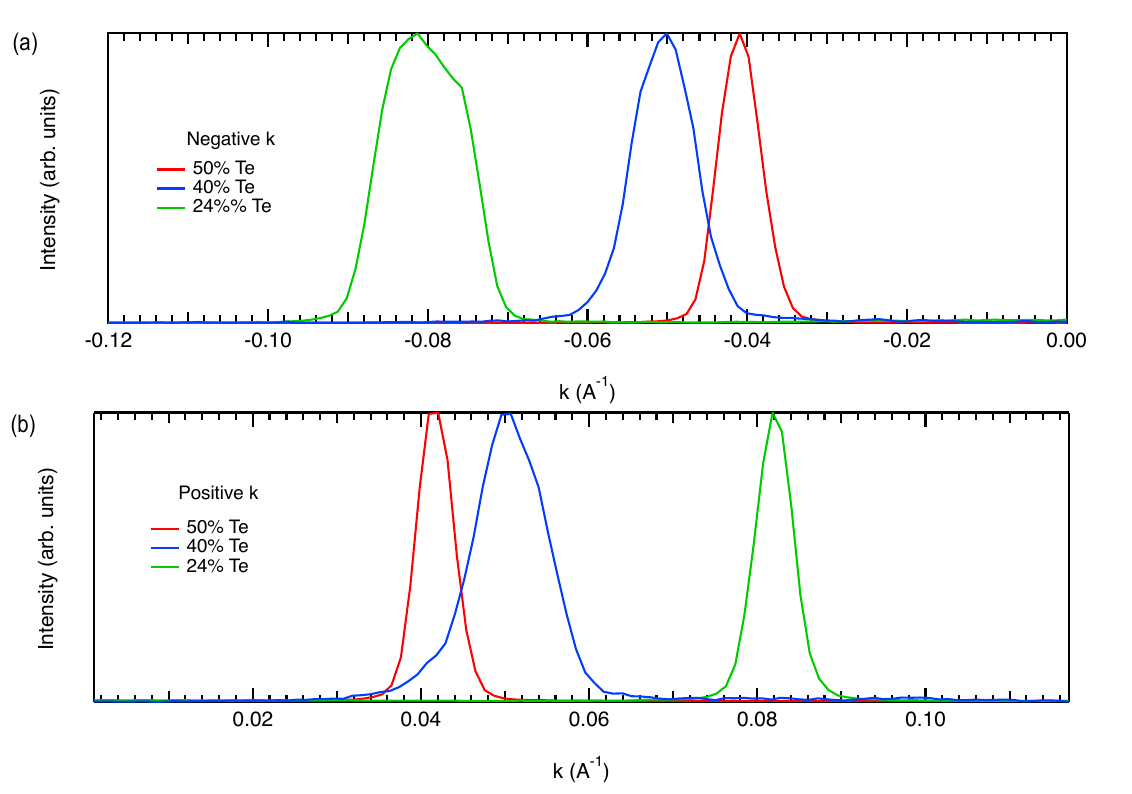}
    \caption{Momentum distribution curves of the surface state bands extracted at the chemical potential. a) for negative momenta. b) for positive momenta.}
\end{figure}
\indent To examine this in more detail, we plot the full width at half maximum (FWHM) of the MDCs as a function of momentum in Figure 5. These plots were obtained by fitting the MDCs with Lorentzian functions. We fit the peaks at positive and negative momentum values to quantify the difference in width on both sides of the $\Gamma$ point. For the medium and high Te concentrations, shown in panels (b) and (c), respectively, we observe a relatively small variation in width as a function of energy. One of the bands is consistently broader than the other; however, the difference is not significant. In the low-Te-concentration sample shown in panel (a), we observe a larger difference in the widths of the left and right bands, with the right band being approximately three times wider than the left. Since all momentum points were measured simultaneously in a single scan, this is unlikely to be a result of measurement non-uniformity. At present, the mechanism behind this phenomenon is not understood, and further theoretical analysis is required to explain it.

\begin{figure}
    \includegraphics[scale=0.43]{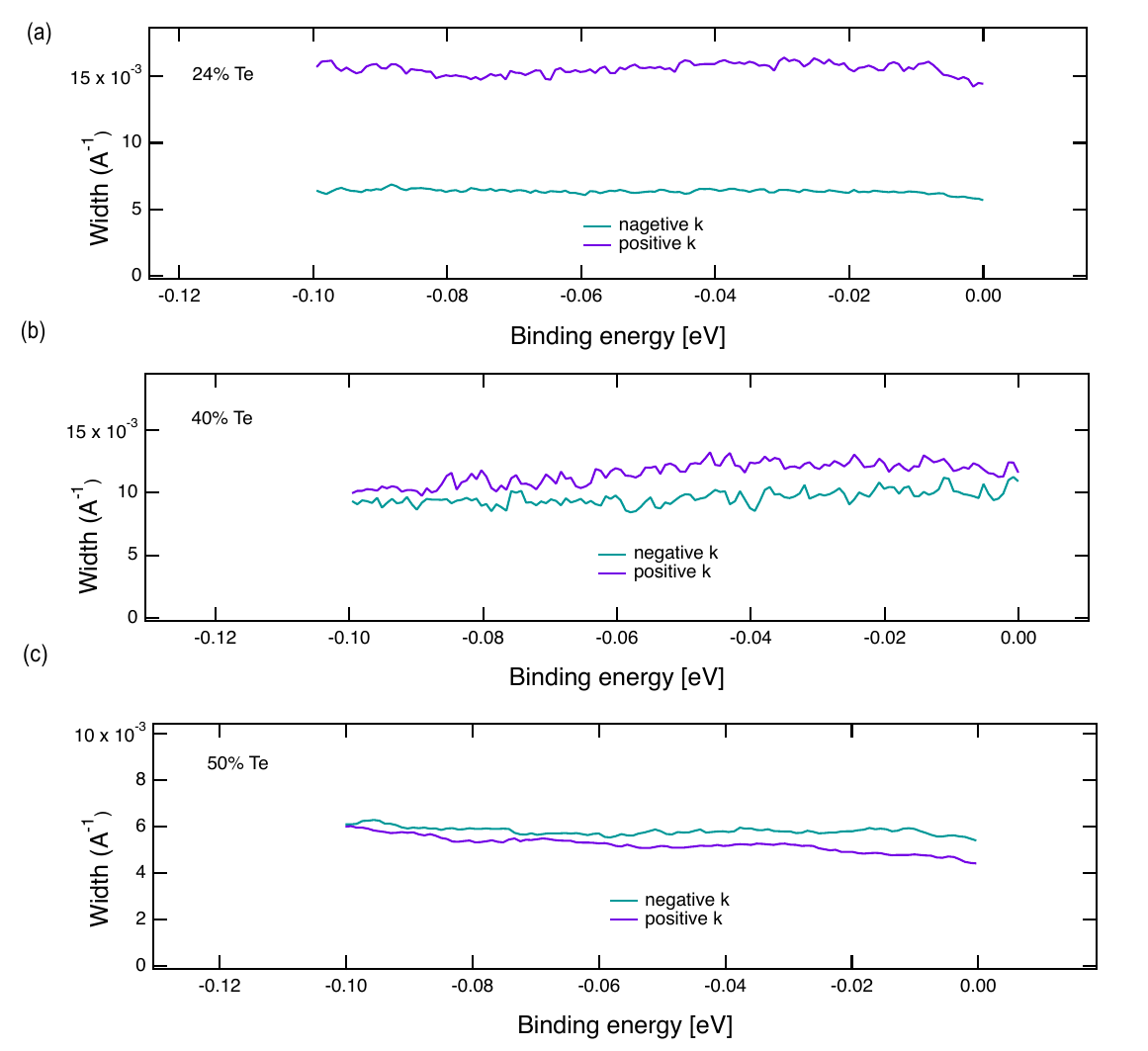}
    \caption{Full-width at half-maximum (FWHM) of the MDC peaks as a function of binding energy, extracted from Lorentzian fits.
    a) for 24~\% Te content, 
    b) for 40~\% Te content, 
    c) for 50~\% Te content} 
    
\end{figure}

\indent One of the key challenges in studying topological insulators and their practical applications is the presence of a bulk band in many stoichiometric compounds. This is clearly visible in Fig. 1(d), where the bulk band crosses the chemical potential even at Te concentrations as high as 24\%. This bulk band makes a significant contribution to the conductivity and prevents the study of the transport properties of the topological surface states. This is illustrated in Fig. 6, where we plot the temperature dependence of the resistivity for three Te concentrations.

For the lowest Te concentration (panel a), where the bulk band dips below the chemical potential, the resistivity exhibits metallic behavior above 150 K, as it increases with temperature. Below 150 K, it displays semiconducting behavior, most likely due to a temperature-induced Lifshitz transition, as also reported in WTe$_2$ \cite{TLifshitz}, which could shift the bottom of the bulk band toward the chemical potential. There is a small but noticeable saturation at very low temperatures, which may indicate a minor contribution to the conductivity from surface states. The overall resistivity remains relatively low ($\sim 4~\Omega$cm), consistent with a dominant contribution from the bulk band.

At 40\% Te, the crossover between metallic and semiconducting behavior shifts to around 230 K, and the low-temperature resistivity increases by two orders of magnitude, indicating a further reduction in the bulk band contribution. Finally, for the highest Te concentration (panel c), the resistivity exhibits semiconducting behavior up to 300 K, decreasing with increasing temperature, and is nearly three orders of magnitude higher at low temperatures than in the low-Te case. At the lowest temperatures, there is a crossover toward saturation, which we interpret as a signature of metallic surface states making a significant contribution to conduction. While the resistivity would be expected to increase with decreasing temperature due to the semiconducting nature of the bulk band, the metallic topological surface states begin to contribute more significantly, slowing the increase in resistivity upon cooling.

We therefore argue that at this Te concentration and at low temperatures, the transport properties are dominated by the topological surface states. This makes Bi$2$(Se${0.5}$,Te$_{0.5}$)$_3$ an excellent candidate for studying the intrinsic transport properties of topological surface states.


\section{\label{sec:level4} Conclusions\protect}
\indent In summary, we presented ARPES studies of electronic structure and temperature dependent resistivity showing how these properties evolve with varying concentrations of Te in \ce{Bi2(Te-Se)3} topological insulator. We observed significant changes in the chemical potential, consistent with decreased electron concentration upon substitution of Se with Te. This results in a decrease of the binding energy of the Dirac point, rounding of the Fermi surface and more insulating behavior of the bulk band. The temperature dependence of the resistivity points to increasingly semiconducting behavior and dominant contribution of the topological surface states at low temperature for the 50~\% Te concentration. These samples appear to be excellent platform for studying topological surfaces states.

\begin{figure}
    \includegraphics[scale=0.4]{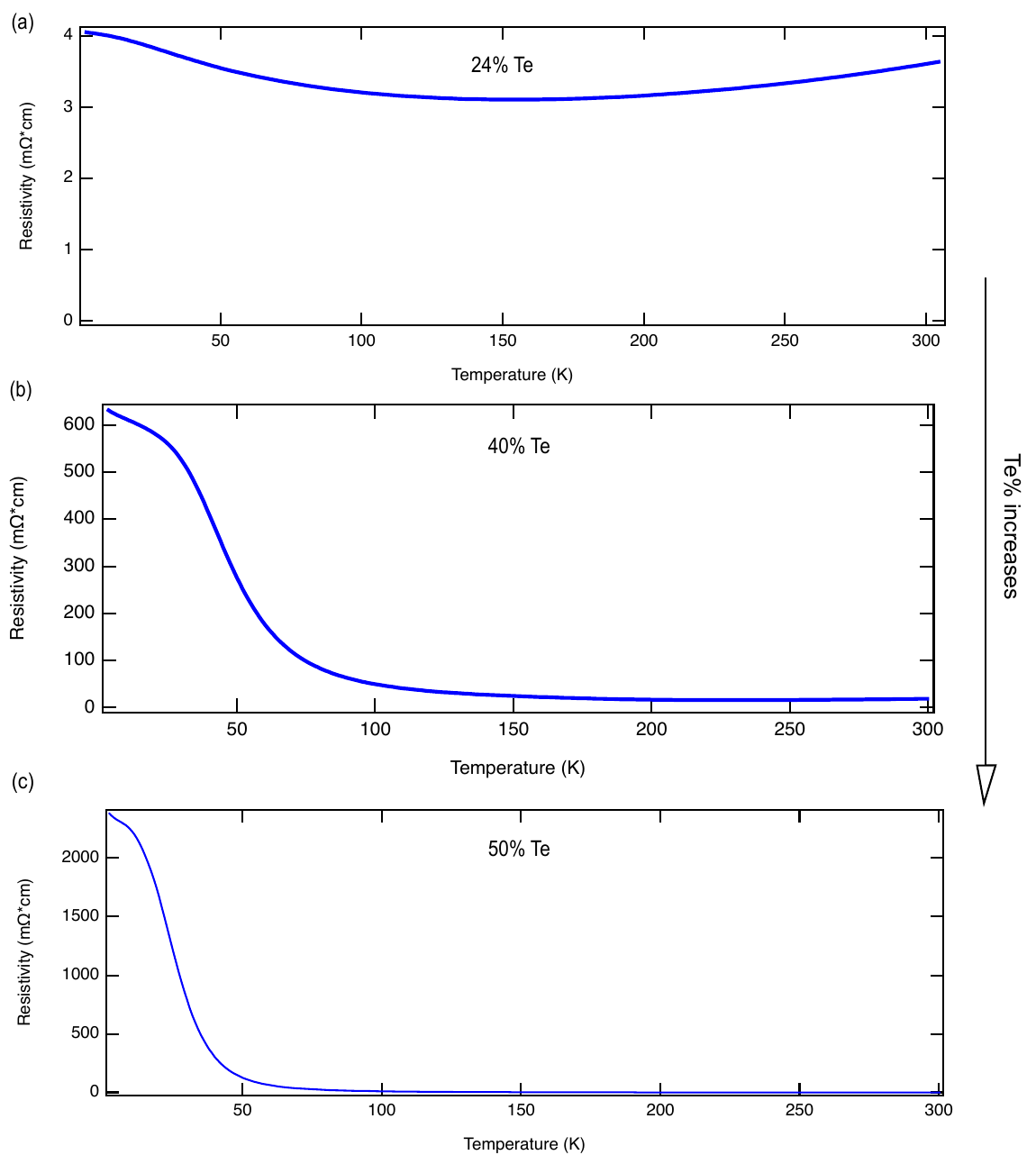}
    \caption{Temperature dependence of the resistivity a) for 24~\% Te content, 
    b) for 40~\% Te content, 
    c) for 50~\% Te content} 
 
\end{figure}


\section{\label{sec:level4} ACKNOWLEDGMENTS \protect} 
\indent This work was supported by the US Department of Energy, Office of Basic Energy Sciences, Division of Materials Science and Engineering. Ames National Laboratory is operated for the US Department of Energy by Iowa State University under Contract No. DE-AC02-07CH11358.

\section{\label{sec:level4} DATA AVAILABILITY\protect} 
The raw data for this paper is available at Harvard Dataverse, DOI: https://doi.org/10.7910/DVN/NR8EMI

\bibliography{refs}

\end{document}